\documentclass[10pt,aps,pra,twocolumn,superscriptaddress,showpacs]{revtex4-1}

\usepackage[pdftex]{graphicx}
\usepackage{hyperref}
\usepackage{amssymb}
\usepackage{amsmath}
\usepackage{xspace}
\usepackage{color}

\newcommand{\trans}{$\leftrightharpoons$}

\usepackage[dvipsnames]{xcolor}

\begin{document} 

\title{Quantum phase transitions in the dimerized extended 
Bose-Hubbard model}

\author{Koudai Sugimoto}
\affiliation{Center for Frontier Science, Chiba University, 
Chiba 263-8522, Japan}

\author{Satoshi Ejima}
\author{Florian Lange}
\author{Holger Fehske}
\affiliation{Institut f{\"u}r Physik, Universit{\"a}t Greifswald, D-17489 Greifswald, Germany}

\date{\today}

\begin{abstract}
We present an unbiased numerical density-matrix renormalization group study of the one-dimensional Bose-Hubbard model supplemented by nearest-neighbor Coulomb interaction and bond dimerization. It places the emphasis on the determination of the ground-state phase diagram and shows that, besides dimerized Mott and density-wave insulating phases, an intermediate symmetry-protected topological Haldane insulator emerges at weak Coulomb interactions for filling factor one, which disappears, however, when the dimerization becomes too large. Analyzing the critical behavior of the model, we prove that the phase boundaries of the Haldane phase to  Mott insulator and density-wave states belong to the Gaussian and Ising universality classes with central charges $c=1$ and $c=1/2$, respectively,  and merge in a tricritical point. Interestingly we can demonstrate a direct Ising quantum phase transition between the dimerized Mott and density-wave phases above the tricritical point. The corresponding transition line terminates at a critical end point that belongs to the universality class of the dilute Ising model with $c=7/10$. At even stronger Coulomb interactions the transition becomes first order.  
\end{abstract}

\maketitle

\section{Introduction}
Over the past years, ultracold atoms in optical lattices have become 
a fascinating tool to explore strongly correlated many-body systems 
and thereby provide also valuable insights into complex phenomena in 
solid-state systems~\cite{RevModPhys.80.885,Bloch12,Gross17}.
Ultracold-atom-based quantum simulators have already been used, e.g.,  to 
observe the transition from a superfluid (SF) to a Mott insulator (MI) phase for bosons~\cite{Greiner02}, 
to realize the crossover between Bose-Einstein condensation and Bardeen-Cooper-Schrieffer pairing~\cite{Zwerger12}, 
or to modulate the range of interactions in quantum systems~\cite{Baier16,Landig16}.

One of the targeted model systems for ultracold atoms
is the Bose-Hubbard model (BHM), which has been intensively studied 
from a theoretical point of view. 
Quite recently, triggered by the observation of a symmetry-protected-topological (SPT) Haldane phase in 
the spin-1 Heisenberg chain~\cite{Ha83,GW09,PTBO10}, the related Haldane insulator (HI) phase 
in the extended BHM (EBHM) with longer-range repulsion~\cite{DBA06} has attracted significant attention.  

Including a bond dimerization, which can also be realized in optical lattices~\cite{Atala13}, the physical properties of the spin-1 chain change drastically, e.g.,
the Haldane phase shrinks rapidly when the dimerization increases and eventually even disappears~\cite{KNO96,KN97}.
In this work, we explore the effect of the bond dimerization $\delta$
in the EBHM using the 
density-matrix renormalization group (DMRG) technique~\cite{White92,Sch11}. 
We especially demonstrate that a direct continuous transition takes place between 
the dimerized MI and density-wave (DW) phases, instead of the first-order transition observed in 
the pure EBHM ($\delta=0$). 

The  paper is structured as follows: Section~\ref{model-method} introduces the EBHM with bond dimerization, as well as   
the numerical techniques for its investigation. The physical 
quantities of interest will be defined in Sec.~\ref{phys-quantities}.
Section~\ref{num-results} presents the ground-state phase diagram of 
the dimerized EBHM for $\rho=1$ and classifies the  phase
boundaries. Some results for band filling factor $\rho=1/2$ can be found in Appendix~\ref{rho-onehalf}. 
Section~\ref{summary} summarizes our results and gives a brief outlook.

\section{Model and Method}
\label{model-method}
As outlined above, we consider the EBHM with an additional explicit bond dimerization $\delta$,
\begin{eqnarray}
 \hat{H}=\hat{H}_{\rm EBHM}
  -t\sum_j\delta(-1)^j
  \left(\hat{b}_{j+1}^\dagger\hat{b}_{j}^{\phantom{\dagger}}
   +{\rm H.c.}
   \right)\,,
\label{model}
\end{eqnarray}
where the EBHM Hamiltonian is given by
\begin{align}
 \hat{H}_{\rm EBHM}=&
  -t\sum_j
  \left(\hat{b}_{j+1}^\dagger\hat{b}_{j}^{\phantom{\dagger}}
   +{\rm H.c.}
   \right)
 +U\sum_j\hat{n}_j\left(\hat{n}_j-1\right)/2
 \nonumber \\
 &  +V\sum_j\hat{n_j}\hat{n}_{j+1}\,.
  \label{ebhm-hamil}
\end{align}
Here, $\hat{b}_j^{\dagger}$ ($\hat{b}_j^{\phantom{\dagger}}$) creates (annihilates) 
a boson at site $j$ of a one-dimensional lattice, and $\hat{n}_j=\hat{b}_j^\dagger\hat{b}_j^{\phantom{\dagger}}$ 
is the corresponding particle number operator. The transfer amplitude $t$ enables the bosons to hop 
between neighboring lattice sites, whereas the  on-site (nearest-neighbor) Coulomb repulsion 
$U$ ($V$) tends to localize the particles by establishing an MI (a DW) ground state, at least when the number 
of bosons $N$ equals the number of lattice sites $L$, i.e., $\rho=N/L=1$. In this case, a finite dimerization 
should also promote an insulating state but now with alternating strong and weak bonds.

The ground-state phase diagram of the pure BHM, where $V=0$ and $\delta=0$, has only two phases, 
an SF and an MI~\cite{KWM00}, which are separated by a Kosterlitz-Thouless phase transition at $t/U\simeq 0.305$ for $\rho=1$~\cite{EFG11}. 
Adding now $V$, and restricting the maximum number of bosons per site $n_{\rm b}$ to be two, the EBHM can be approximately mapped onto the spin-1 $XXZ$ model with single-ion anisotropy, whereby the bosonic operators  $\hat{b}_j^\dagger$, $\hat{b}_j^{\phantom{\dagger}}$, and $\hat{n}_j$ will be replaced by the spin-1 operators $\hat{S}_j^+$, $\hat{S}_j^-$,  and $\hat{S}_j^z+1$, respectively~\cite{AA02}. 
As a result, an SPT Haldane insulator  appears between the MI and DW phases for intermediate couplings~\cite{DBA06,BDGA08},
which resembles the gapped Haldane phase of the quantum spin-1 Heisenberg chain~\cite{Ha83}. We note that the HI phase continues to exist if one includes 
higher boson numbers $n_{\rm b}>2$~\cite{ELF14,EF15}. In the DMRG calculations, a finite maximum number of bosons per site $n_{\rm b}$ must be used. All results for $\rho=1$ in the main text are obtained with $n_{\rm b}=4$. 

To explore the effects of the dimerization in the full model~\eqref{model}, 
we employ the matrix-product-state-based infinite DMRG (iDMRG) technique~\cite{Mc08}. The iDMRG 
provides us with unbiased numerical data directly in the thermodynamic limit.  Hence the 
phase boundaries can be obtained without any finite-size scaling procedure.   On the other hand, 
we determine the critical behavior by tracking the central charge along the quantum phase transition (QPT) lines 
through the use of the more standard DMRG technique for finite systems with periodic boundary conditions (PBC). 
The quantum phase transition itself is characterized  by various excitation gaps obtained by combining DMRG 
and infinite matrix-product-state representation at the boundaries of the system~\cite{Phien12,LEF15}

\section{System characterization}
\label{phys-quantities}
Now we present the physical quantities of interest and  explain how they can be simulated within the (i)DMRG framework.

\subsection{Entanglement spectrum, central charge and correlation length}
To determine SPT states in the model~\eqref{model}, we discuss 
the so-called entanglement spectrum $\varepsilon_\alpha$~\cite{PhysRevLett.101.010504}, 
which can be extracted from the Schmidt decomposition. 
Dividing the system with $L$ sites into two subblocks, 
${\cal H}={\cal H}_{\ell}\otimes{\cal H}_{L-\ell}$, 
and considering the reduced density matrix $\rho_{\ell}={\rm Tr}_{L-\ell}[\rho]$
of a sub-block of (arbitrary) length $\ell<L$, the entanglement spectrum is given by the singular values
$\lambda_\alpha$ of $\rho_{\ell}$ as $\varepsilon_\alpha=-2\ln\lambda_\alpha$.
If we split the system into two semi-infinite pieces during the iDMRG simulations, 
the entanglement levels $\varepsilon_\alpha$ show a characteristic degeneracy in the SPT phase, 
as has been demonstrated for the Haldane phase of the spin-1 chain~\cite{PTBO10}. 

The entanglement spectrum also yields valuable information about the criticality of the system. 
For the von Neumann entanglement entropy, $S_L(\ell)=-\sum_\alpha \lambda_\alpha^2\ln \lambda_\alpha^2$, field theory predicts that
\begin{eqnarray}
 S_{L}(\ell)=
  \frac{c}{3}\ln\left[\frac{L}{\pi}\sin\left(\frac{\pi\ell}{L}\right)\right]
  +s_1
  \label{ee}
\end{eqnarray}
in a critical system with PBC~\cite{CC04}. In Eq.~\eqref{ee}, $c$ is the central charge and $s_1$ is a nonuniversal constant. 
Employing a doubled unit cell,  in view of the explicit dimerization, the central charge can be calculated 
very efficiently from the relation~\cite{Ni11}
\begin{eqnarray}
 c^\ast(L)=\frac{3[S_L(L/2-2)-S_L(L/2)]}{\ln\{\cos[\pi/(L/2)]\}}\,.
 \label{cstar}
\end{eqnarray}

In addition, within an iDMRG calculation, the correlation length $\xi_\chi$ can be obtained from the second 
largest eigenvalue of the transfer matrix for some bond dimension $\chi$~\cite{Mc08,Sch11}. While the physical 
correlation length diverges when the system becomes critical, $\xi_\chi$
stays finite during the numerical simulations due to the finite bond dimension. 
Nevertheless, $\xi_\chi$ can be utilized to determine the phase transition point 
because it develops a pronounced maximum with increasing $\chi$ 
near the critical point. Putting these criteria together, the QPT can be determined 
with high precision. 

\subsection{Excitation gaps}
To determine the criticality of the QPTs one can simulate various excitation gaps of the model~(\ref{model}), just as for the EBHM~\cite{DBA06,BDGA08,ELF14}.
 For instance, in the EBHM, the single-particle gap 
\begin{eqnarray}
 \Delta_{\rm sp}&=&E_0(N+1)+E_0(N-1)-2E_0(N)
 \label{particle-gap}
\end{eqnarray}
closes at the MI-HI transition, and the neutral gap 
\begin{eqnarray}
 \Delta_{\rm n}&=&E_1(N)-E_0(N)
 \label{neutral-gap}
\end{eqnarray}
vanishes at the MI-HI and HI-DW transitions, where 
$\Delta_{\rm n}$ closes linearly in the latter case, 
indicating a critical exponent $\nu=1$ of the Ising universality class. 
In Eqs.~\eqref{particle-gap} and \eqref{neutral-gap}, 
$E_0$ ($E_1$) denotes the ground-state energy (energy of the first excited state) of the finite $L$-site system
with fixed boson number.

\subsection{Density-wave order parameter}
By analogy with the charge-density-wave order parameter of the fermionic Hubbard-type 
models~\cite{EELF16,ELEF18}, a (dimerized) DW state in the model~\eqref{model} can be characterized by 
a nonvanishing expectation value of the operator  
\begin{eqnarray}
 \hat{m}_{\rm DW}=\frac{1}{L}\sum_j(-1)^j(\hat{n}_j-1)\,.
\end{eqnarray}
Most importantly,  analyzing $\langle\hat{m}_{\rm DW}\rangle$ close to the Ising or the tricritical Ising transitions points provides 
the critical exponent $\beta$~\cite{ELEF18}.

\section{Numerical results for $\boldsymbol{\rho=1}$}
\label{num-results}
\subsection{Ground-state phase diagram}
\begin{figure}[tb]
 \includegraphics[width=0.99\columnwidth]{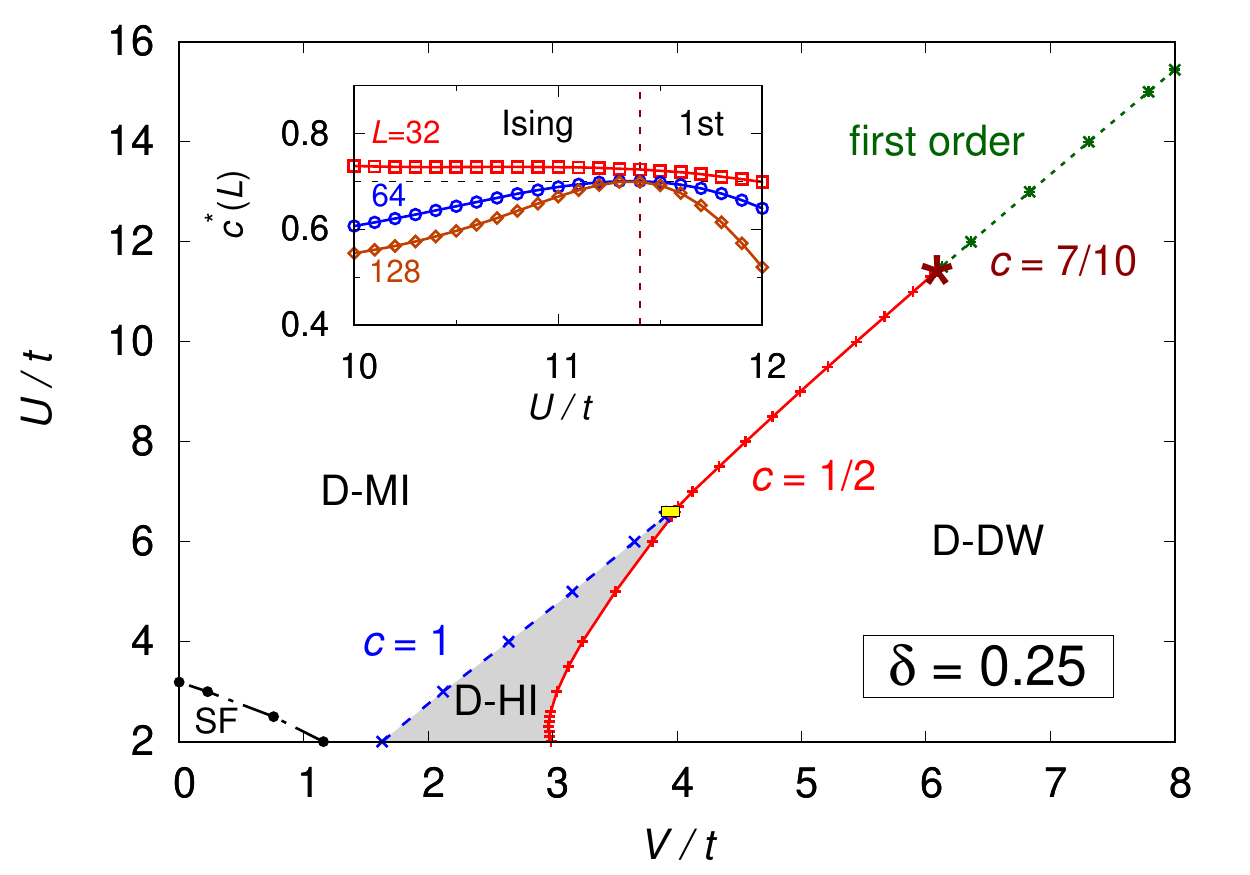}
 \caption{IDMRG ground-state phase diagram of the dimerized EBHM~\eqref{model} 
 for $U/t\geq 2$ with $\delta=0.25$ and $n_{\rm b}=4$. 
 Here the blue dashed line gives the D-MI$\leftrightharpoons$D-HI phase 
 boundary; the red solid line denotes the continuous Ising phase transition. 
 Both lines merge at the tricritical point located inside the small rectangle which is enlarged in 
 Fig.~\ref{tricritical-pd}(c) (see also the discussions in the text). 
 The QPT is continuous (first order) below (above) the critical end point ($V/t, U/t)_{\rm ce}$ marked by the star symbol 
 [there we obtain for the central charge $c^\ast(L)\simeq 0.7$  from Eq.~\eqref{cstar}  as $L\to\infty$ 
 on the D-MI\trans D-DW transition line, see inset]. 
 In the weak ($V,U$)-coupling regime an SF phase is formed.
 }
 \label{pd1}
\end{figure}

Figure~\ref{pd1} presents the  ground-state phase diagram of the EBHM 
with an explicit bond dimerization $\delta=0.25$ and $n_{\rm b}=4$ obtained by 
iDMRG. For the considered weak dimerization, we observe, just as for the 
EBHM  ($\delta=0$), an HI between the MI and DW states, but now these phases 
exhibit a finite bond dimerization, i.e., actually we have D-HI, D-MI, and D-DW states. 
For weak onsite and nearest-neighbor repulsions, an SF phase appears. 
Additionally, there may be a region of phase separation for $U/t<2$,
as observed in the model without dimerization~\cite{PhysRevB.90.205123}. 
Here, however, we restrict ourselves to the parameter regime $U/t\geq 2$ 
in order to concentrate on the study of D-MI, D-HI, and D-DW phases 
and the transitions between them. 

Also the universality classes of the QPT between the D-HI and the D-MI
(D-HI and D-DW) phases are the same as for the EBHM, where they are characterized by 
a central charge $c=1$ ($c=1/2$). The relevant difference is that now the transition between the D-MI and 
D-DW phases is continuous below a critical end point $(V/t, U/t)_{\rm ce}$   [which roughly is (11.4, 6.08) for $\delta=0.25$]. 
The continuous transition also belongs to the Ising universality class, except for the critical end point, which belongs 
to the universality class of the dilute Ising model with $c=7/10$. This will be confirmed numerically 
below.

\subsection{D-HI\trans D-MI and D-HI\trans D-DW quantum phase transitions}
\begin{figure}[tb]
 \includegraphics[width=0.95\columnwidth]{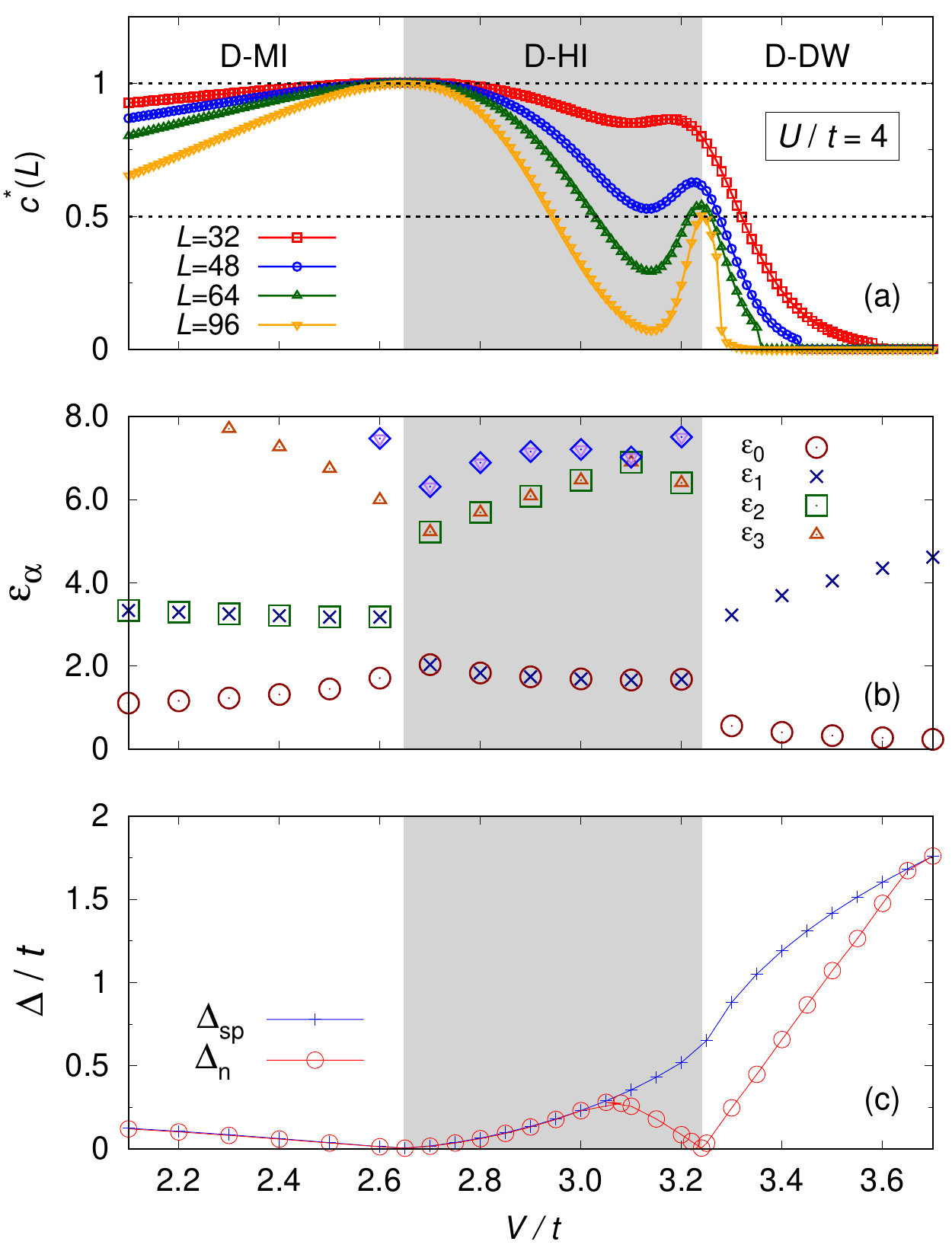}
 \caption{
 Central charge (a), entanglement spectrum (b) and excitation gaps (c) 
 of the dimerized EBHM~\eqref{model} as a function of $V/t$ 
 at fixed $U/t=4$, where $\rho=1$ and $n_{\rm b}=4$. A central charge $c=1$ ($c=1/2$)
 indicates the D-HI\trans D-MI (D-HI\trans D-DW) transition. The D-HI phase is marked in gray.   
 }
 \label{u4}
\end{figure}

We now investigate the nature of the SPT D-HI state and its phase boundaries
in more detail. Figure~\ref{u4}(a) displays the behavior of the central 
charge $c^\ast(L)$ as a function of $V/t$ at fixed $U/t=4$, which is obtained by evaluating Eq.~\eqref{cstar} by 
DMRG for up to $L=96$ sites with PBC. Increasing the system size 
two peaks develop, which indicates the D-MI\trans D-HI and D-HI\trans D-DW transitions. 
For $L=96$, we find $c^\ast\simeq 1.000$ ($c^\ast\simeq 0.503$) 
at $V_{\rm c1}/t\simeq 2.65$ ($V_{\rm c2}/t\simeq 3.24$), which points toward 
a Gaussian (an Ising) QPT. The corresponding entanglement spectrum 
$\varepsilon_\alpha$ [Fig.~\ref{u4}(b)] underlines that a nontrivial topological phase is 
realized for $V_{\rm c1}< V< V_{\rm c2}$,  because the entanglement 
levels show  the characteristic degeneracy demonstrated previously  
for the Haldane phase of the spin-1 chain~\cite{PTBO10}.

Figure~\ref{u4}(c) clearly shows the  different behavior 
of  the excitation gaps in the diverse insulator phases, as well as  
at their phase boundaries:  The single-particle gap $\Delta_{\rm sp}$ is finite 
throughout the phase diagram, except for the D-HI\trans D-MI QPT, 
whereas the neutral gap $\Delta_{\rm n}$ closes both at the D-MI\trans D-HI and D-HI\trans D-DW QPTs. 
At the D-HI\trans D-DW transition $\Delta_{\rm n}$ closes linearly, which reflects the critical exponent $\nu=1$ of the Ising universality class. 
Nevertheless, the D-HI phase and its phase boundaries display the same behavior as for the nondimerized EBHM. 
Note that the D-HI phase disappears at the tricritical point $(V/t, U/t)_{\rm tr}$ [which is located at (4.1,6.9) for $\delta=0.25$], where the central
charge becomes 1.

\subsection{D-MI\trans D-DW Ising transition}
\begin{figure}[tb]
 \includegraphics[width=0.95\columnwidth]{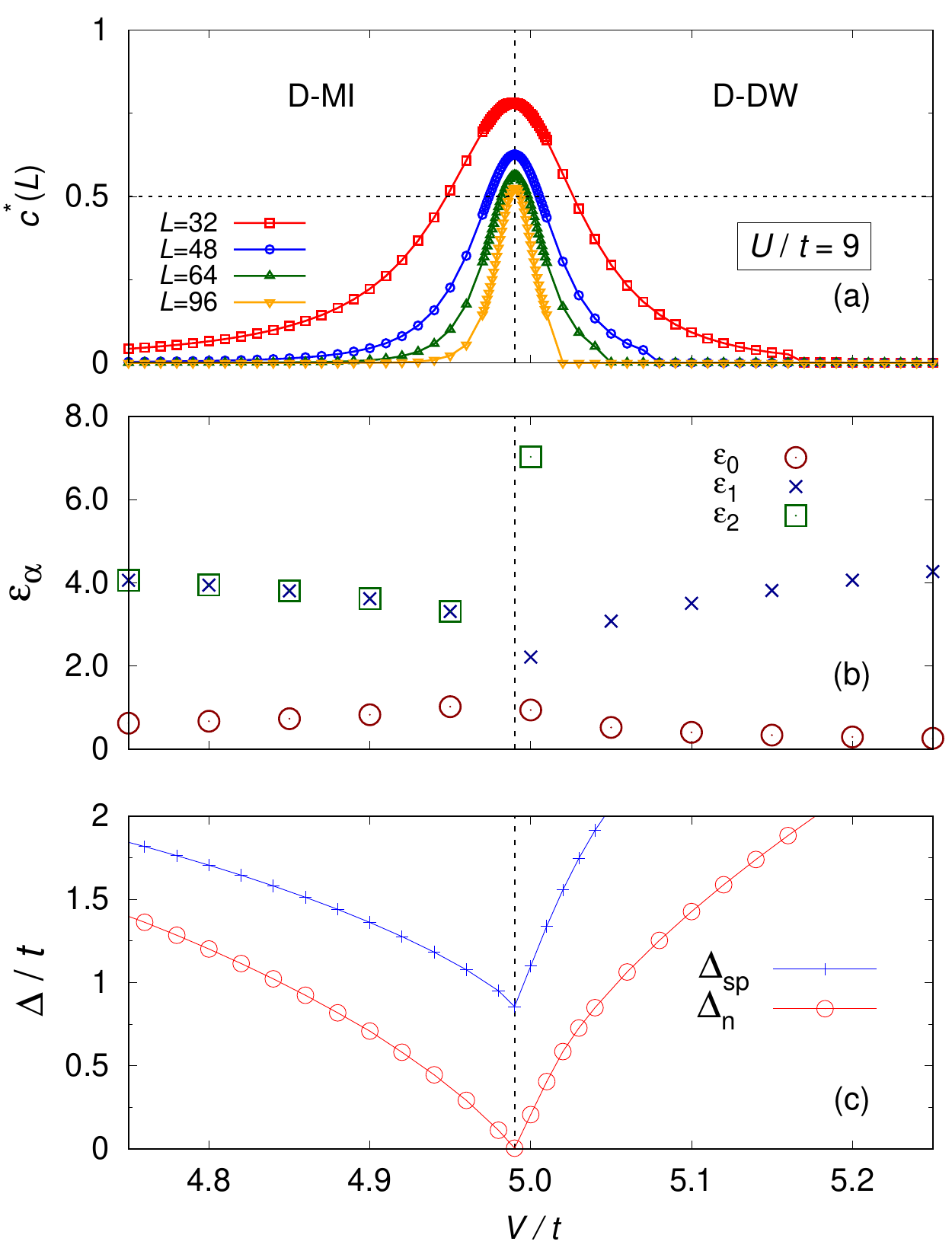}
 \caption{
 Central charge (a), entanglement spectrum (b), and excitation gaps (c) 
 of the model~\eqref{model} as a function of $V/t$ 
 for fixed $U/t=9$ with $\rho=1$ and $n_{\rm b}=4$. 
 The data indicate a D-MI\trans D-DW transition with $c=1/2$.
 }
 \label{u9}
\end{figure}

The most significant effect of the dimerization is the direct Ising transition 
between the D-MI and D-DW phases which could not be observed in the pure EBHM. 
Figure~\ref{u9}(a) displays the central charge 
$c^\ast(L)$,  obtained from Eq.~\eqref{cstar} by DMRG. 
Obviously, in the vicinity of the D-MI\trans D-DW transition, a peak develops
which gets sharper if the system size $L$ is increased. Fixing $U/t=9$, 
we find $c^\ast\simeq 0.526$ at $V_{\rm c}\simeq 4.99$, indicating that the QPT 
belongs to the Ising universality class. Since the D-HI phase is absent,  the entanglement 
spectrum $\varepsilon_\alpha$ is no longer degenerate [in the remaining D-MI and D-DW phases,
cf. Fig.~\ref{u9}(b)].  Figure~\ref{u9}(c) gives the excitation gaps
for $U/t=9$. Again the single-particle gap $\Delta_{\rm sp}$ stays finite,
and the neutral gap $\Delta_{\rm n}$ closes at the D-MI\trans D-DW transition point 
linearly, i.e., $\nu=1$ (Ising universality class). 

As already pointed out, the continuous Ising transition line between 
D-MI and D-DW phases terminates at the tricritical Ising transition point. 
The inset of Fig.~\ref{pd1} shows how a pronounced maximum develops
in the  central charge $c^\ast$ on the D-MI\trans D-DW 
transition line as $L$ increases. We obtain $c^\ast\simeq 0.699$ at the critical end point
$(V/t, U/t)_{\rm ce}\simeq (6.083,11.4)$, in agreement with the prediction of 
field theory for the universality class of the dilute Ising 
model, $c=7/10$.

\subsection{Tricritical regime}
\begin{figure}[tb]
 \includegraphics[width=0.95\columnwidth]{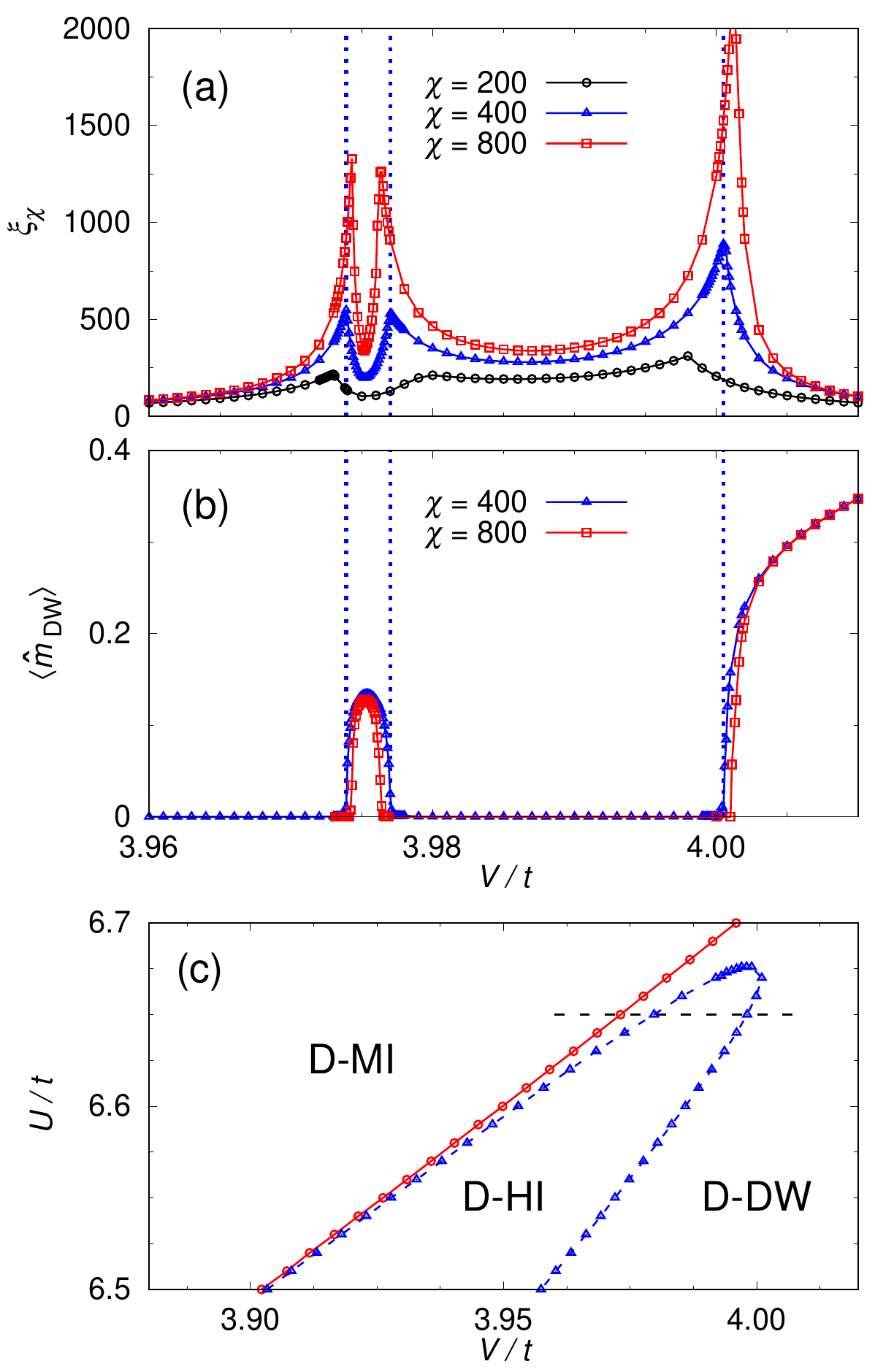}
 \caption{
 (a) Three-peak structure of the correlation length $\xi_\chi$ of the dimerized EBHM with $\delta=0.25$ and $U/t=6.65$. $\chi$ gives the bond dimension used in iDMRG. 
 (b) Corresponding behavior of the DW order parameter $\langle \hat{m}_{\rm DW}\rangle$. Note that $\langle \hat{m}_{\rm DW}\rangle$ is finite not only for $V/t\gtrsim 4.00$
 but also for $3.974\lesssim V/t \lesssim 3.977$. The dotted lines denote the QPT points with $\chi=400$. 
 (c) Zoomed-in phase diagram in the immediate vicinity of the  tricritical point. The dashed line
 illustrates the parameter scan performed in panels (a) and (b). Note that the parameter region of panel (c) is equal to 
 the size of the rectangle in Fig.~\ref{pd1}. 
 }
 \label{tricritical-pd}
\end{figure}

To investigate the surroundings of the tricritical point where the D-HI phase vanishes, 
and determine the value  of  $(V/t, U/t)_{\rm tr}$ with maximum precision,
we calculated the correlation length $\xi_\chi$ varying $V/t$, at fixed $U/t$, above and below
the tricritical point. Here a single-peak, respectively, two-peak structure, would be expected.
From Fig.~\ref{tricritical-pd}(a) it seems, however, that in the immediate vicinity 
of the tricritical point a three-peak structure appears.
That is, the DW order parameter $\langle\hat{m}_{\rm DW}\rangle$ becomes
finite not only for $V>V_{\rm c3}$ but also for $V_{\rm c1}<V<V_{\rm c2}$ 
[see Fig.~\ref{tricritical-pd}(b)], where $V_{\rm c1}<V_{\rm c2}<V_{\rm c3}$ 
denote the positions of three peaks. Plotting the position of these peaks when $U/t$ is changed,
we obtain the strongly zoomed-in phase diagram depicted in Fig.~\ref{tricritical-pd}(c).  
According to this figure, the D-DW phase penetrates between the D-MI and the D-HI phase near the tricritical point $(V/t, U/t)_{\rm tr}$.  Since this re-entrance behavior of the D-DW phase is found  numerically in a very limited parameter range only, and $V_{\rm c2}$ still shifts in the direction of  $V_{\rm c1}$ as $\chi$ increases [see~Fig.~\ref{tricritical-pd}(a)],  it would be highly desirable to explore this region or behavior more thoroughly, e.g., accompanying our iDMRG calculations by field theory, which is beyond the scope of this work, however.

\section{Summary and conclusions}
\label{summary}
In this work we explored the ground-state phase diagram of the extended Bose-Hubbard model with bond dimerization for filling factor $\rho=1$ by means of various density-matrix renormalization group techniques. Most notably, we prove the existence a of a symmetry-protected-topological (dimerized) HI which separates---at sufficiently weak Coulomb interactions and dimerization---MI and DW states. In addition, we demonstrate a direct Ising transition line between the MI and DW phases for larger Coulomb interactions, which terminates at a tricritical Ising point (end point) with central charge $c=7/10$, where it becomes first order.

The phase diagram of the nondimerized model for $\rho=1$ can be understood by analogy to the spin-1 $XXZ$ chain 
with single-ion anisotropy, with the MI, HI, and DW phases corresponding to the large-$D$, Haldane, and N\'eel phases, respectively. 
In particular, it follows that the HI phase is a symmetry-protected-topological phase, which is protected 
by a modified bond-centered inversion symmetry~\cite{PTBO10}. 
Since this symmetry is respected by the explicit dimerization, the distinction between MI and HI survives in the dimerized model. 
For weak Coulomb repulsions $U$ and $V$, the system realizes an SF phase, 
just as for filling factor $\rho=1/2$, where no MI exists at all in the absence of dimerization. 
If the onsite repulsion $U$ is sufficiently large in the latter case, 
adding a small bond dimerization opens an energy gap so that the system passes into a symmetry-protected-topological dimerized MI phase (see Appendix~\ref{rho-onehalf}). 

We wish to stress that it is extremely difficult to obtain numerical results with sufficient accuracy in the immediate vicinity of the tricritical point. In consequence, it remains an open question whether the observed intervening dimerized DW will survive the limit of infinite bond dimensions in the infinite density-matrix renormalization group simulation, or the tricritical point will be simply shifted to somewhat greater values of the Coulomb interactions. In order to clarify this issue, an elaborate  bosonization-based  field theory would be very helpful. Recently, a field theory analysis was carried out in the dimerized spin-1 $XXZ$ chain~\cite{SciPost}, where the re-entrance behavior of the dimerized N\'eel phase might also occur.

Equally interesting would be an experimental realization of the dimerized extended Bose-Hubbard model by ultracold atomic gases in optical lattices in order to prove or disprove our theoretical predictions regarding the criticality and nontrivial topological properties.

\section*{Acknowledgments}
We thank T. Yamaguchi for fruitful discussions. 
K.S. is grateful for the hospitality at the University of Greifswald. 
F.L. was supported by Deutsche Forschungsgemeinschaft (Germany) through Project No. FE 398/8-1. 
The iDMRG simulations were performed using the ITensor library~\cite{ITensor}.

\appendix
\section{Case $\boldsymbol{\rho=1/2}$}
\label{rho-onehalf}
At vanishing dimerization and a boson filling factor $\rho=1/2$,  a Kosterlitz-Thouless 
transition occurs between the SF and DW phases, in close 
analogy to the metal-insulator transition of the fermionic extended Hubbard model at quarter filling~\cite{MZ93,EGN05}. 
At finite bond dimerization $\delta$ one expects that the SF phase 
gives way to an SPT  D-MI phase~\cite{PhysRevLett.110.260405}. 
Then a continuous Ising
phase transition might occur between the SPT D-MI and the D-DW (just as in the charge sector of the quarter-filled extended Hubbard model 
with explicit dimerization~\cite{TO02,EGN06}).  It is well known that the model~\eqref{model} with $\rho=1/2$ can be mapped onto 
the spin-1/2 dimerized $XXZ$ model if we take the limit $U\gg t$, $V$ and consider only the two lowest Fock states  per site $|0\rangle$ and $|1\rangle$.  
In this case, one may replace $\hat{b}_j^\dagger$, $\hat{b}_j^{\phantom{\dagger}}$, and $\hat{n}_j$
by spin-1/2 operators $\hat{S}_j^+$, $\hat{S}_j^-$, 
and $\hat{S}_j^z+1/2$, respectively~\cite{AA98}, so that the Hamiltonian~\eqref{model} becomes
\begin{align}
\hat{H}= &-t\sum_j [1 + \delta (-1)^j]( \hat{S}_j^+ \hat{S}_{j+1}^- + \hat{S}_j^- \hat{S}_{j+1}^+)  \nonumber \\  &+ V  \sum_j \hat{S}_j^z \hat{S}_{j+1}^z \, . 
\label{xxz-delta}
\end{align}
By taking this limit and $\delta \to \pm 1$, the ground state in the D-MI phase can be adiabatically connected to a fully dimerized state with ``singlets'' at every second bond, which implies that the D-MI is an SPT phase protected by inversion symmetry about the strong bonds. 
This is in contrast to the D-HI for $\rho=1$, which is protected by inversion about both strong and weak bonds, and the D-MI for $\rho=1$, which is a topologically trivial phase.

\begin{figure}[t]
 \includegraphics[width=0.95\columnwidth]{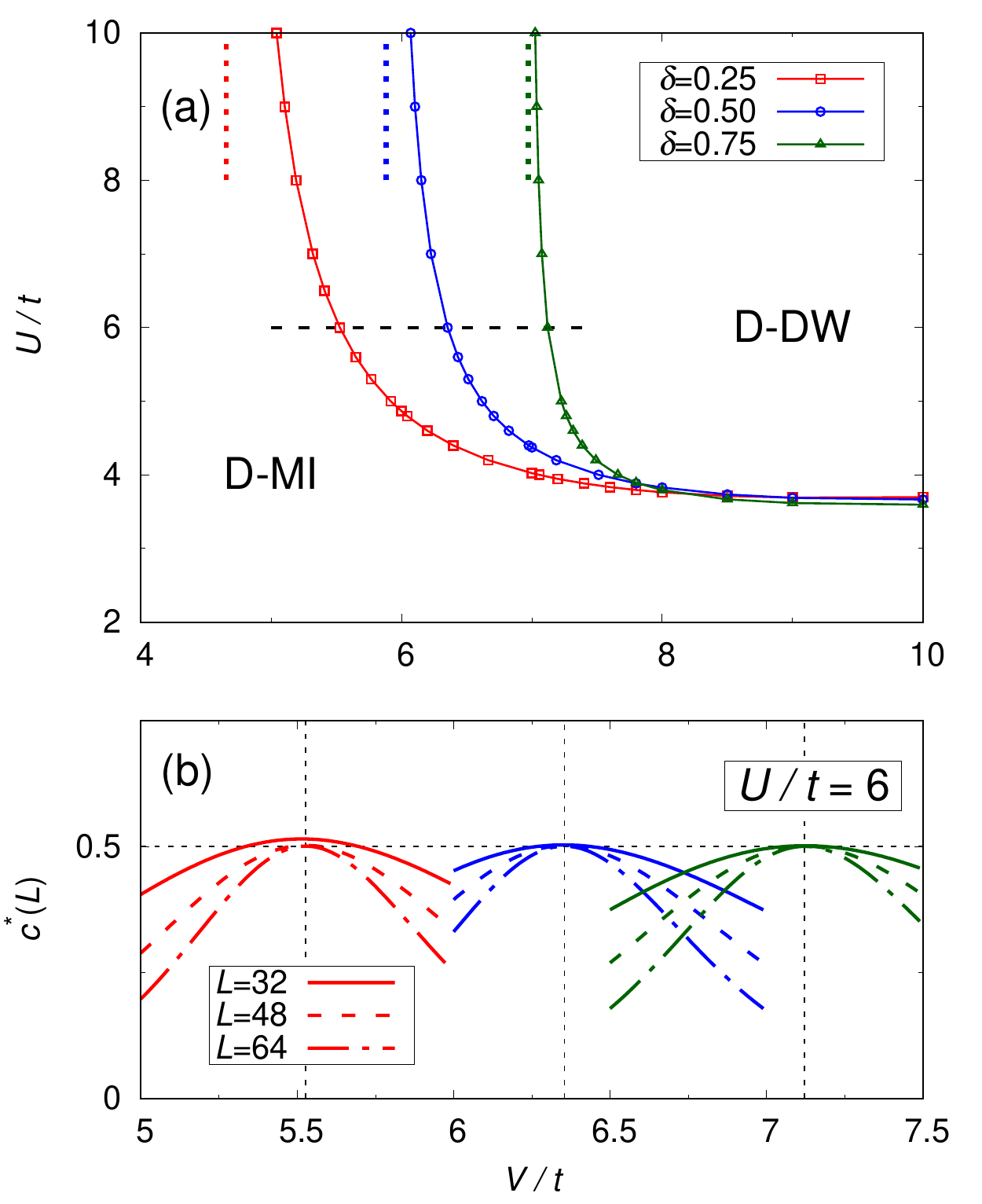}
 \caption{(a) Ground-state phase diagram of 
 the dimerized EBHM with  $\rho=1/2$ and $n_{\rm b}=2$. 
 Data obtained by iDMRG. The dotted lines denote the QPT point
 with same value of $\delta$ in the spin-1/2 chain~\eqref{xxz-delta}.  
 (b) Central charge $c^\ast(L)$ as a function of $V/t$ at fixed $U/t=6$, 
 calculated [along the dashed line in panel (a)] by finite-system DMRG 
 with PBC. 
 }
 \label{pdhalf}
\end{figure}

Figure~\ref{pdhalf}(a) displays the ground-state phase diagram 
for a maximum number of bosons per site $n_{\rm b}=2$ and different bond dimerizations  $\delta=0.25$, $0.50$, and $0.75$.  Only D-MI and D-DW
phases appear.  
The phase boundaries for different $\delta$ approximately coincide for strong nearest-neighbor interactions $V/t>8$. 
In the limit $V\to \infty$, the ground state in the D-DW phase becomes a product state with alternating empty and single-occupied sites. 
The lowest-lying excited state then consists of a single double-occupied site with energy $U$ and two domain walls with energies $-2(t+\delta)$ and $-2(t-\delta)$. Accordingly, the D-DW state should break down at $U/t=4$ for all dimerizations. For smaller $U/t$, phase separation should occur since the D-MI phase is prohibited by the strong nearest-neighbor repulsion. The critical value $U/t=4$ roughly agrees with our numerical results for $V/t \lesssim 10$. However, in the parameter region studied, the D-DW borders only on the D-MI and no phase separation is observed. 

The universality class of the QPT between the D-MI and the D-DW is deduced from the central charge $c^\ast(L)$ [Eq.~\eqref{cstar}] 
by DMRG with PBC.  The observed value $c^\ast\simeq 0.5$ indicates that
the transition belongs to the Ising universality class in two dimensions.

\begin{figure}[tb]
 \includegraphics[width=0.95\columnwidth]{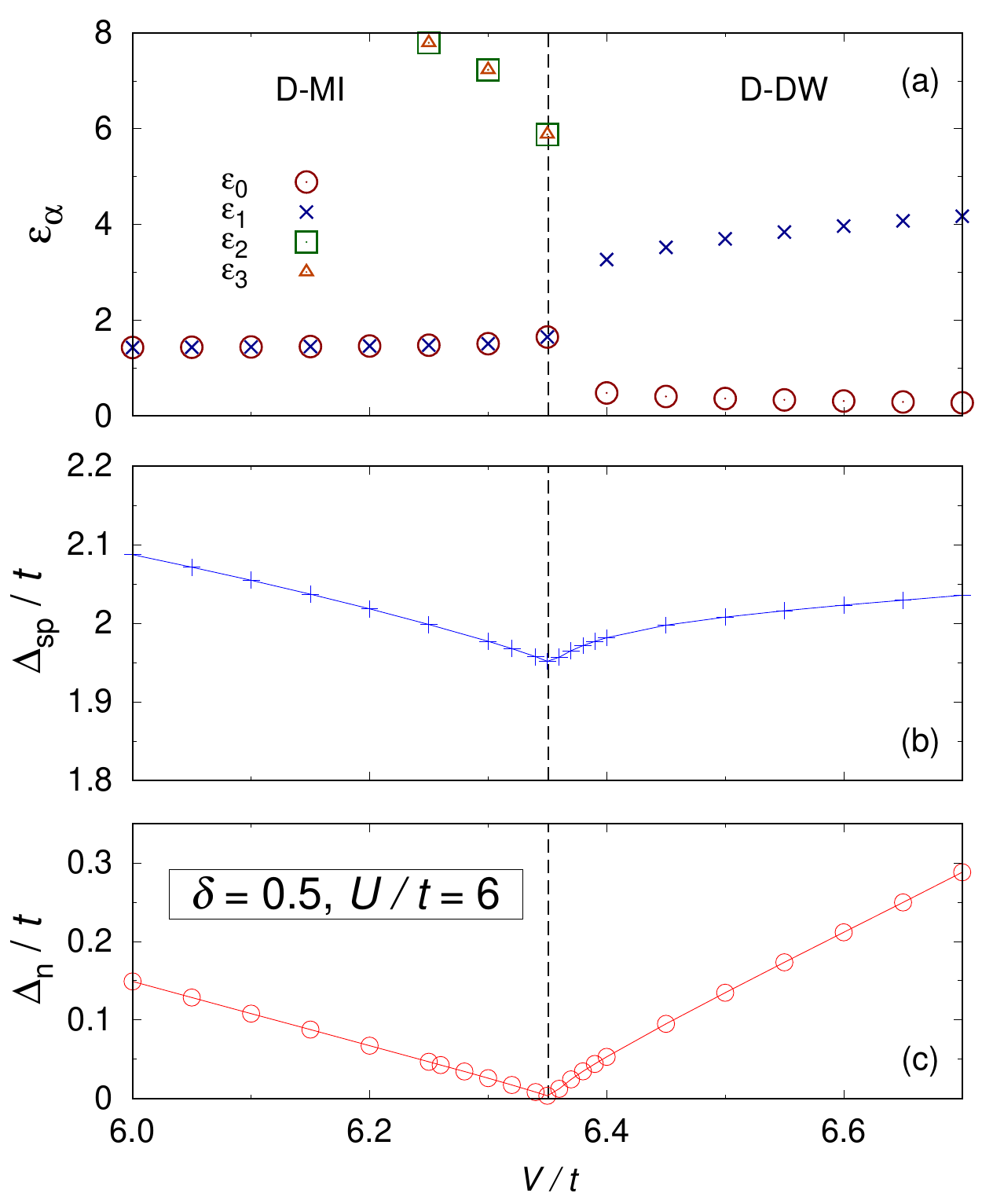}
 \caption{Entanglement spectrum (a), single-particle gap (b),
 and neutral gap (c) in the dimerized EBHM with $\delta=0.5$
 and $n_{\rm b}=2$ at $U/t=6$. The dashed line marks the Ising QPT point at 
 $V_{\rm c}/t\simeq 6.351$.
 }
 \label{gap-half}
\end{figure}

Other static properties of the dimerized EBHM are given by  Fig.~\ref{gap-half} 
for a bond dimerization $\delta=0.5$ and $U/t=6$. Since the D-MI with doubled unit cell
is a nontrivial SPT phase, the D-MI entanglement spectrum exhibits the characteristic 
degeneracy, which is lifted in the D-DW phase. Figure~\ref{gap-half}(b) gives 
the single-particle gap for the same parameter set, which has a minimum at the Ising transition point. 
As in the case of $\rho=1$,  the neutral gap $\Delta_{\rm n}$ closes linearly at the Ising transition point [see Fig.~\ref{gap-half}(c)], yielding 
the critical exponent $\nu=1$ of the Ising universality class.

%


\end{document}